\documentstyle[aps,prb]{revtex}
\begin{document}
\title{Internal transitions in the confined biexciton.}
\author{Ricardo P\'{e}rez$^{1,2}$\cite{ricardo} and 
 Augusto Gonzalez$^{1,2}$\cite{augusto}}
\address{$^1$Instituto de Cibernetica, Matematica y Fisica,
 Calle E 309, Vedado, Ciudad Habana, Cuba.\\
 $^2$Departamento de Fisica, Universidad de Antioquia, A. A. 1226,
 Medellin, Colombia.}
\date{\today}
\maketitle

\begin{abstract}
Optical internal transitions relevant to ODR experiments are computed 
for the biexciton in a quantum dot in the strong confinement regime.
Valence sub-band mixing effects are taken into account in second order 
pertubation theory. The transition probability from the ground state
is concentrated in a few states with relatively high excitation
energies. Level collissions with oscillator strength transfer are
observed as the magnetic field is raised.

\end{abstract}

\vspace{1cm}
The Optical Detection of far-infrared Resonances \cite{odr} (ODR)
have proven to be an efficient method in the study of a sector of the
exciton spectrum not accesible to inter-band transitions. A 
far-infrared radiation, in resonance with an internal transition,
induces changes in the population of the exciton ground-state, 
changing in this way the amplitude of the luminescence peak. 
Usually, in order to tune into resonance a particular transition,
a magnetic field is used instead of varying the frequency of the
infrared source. By this means, many of the exciton transitions
as well as electron and hole cyclotron resonance transitions have
been clearly observed. In addition to this important result, the two 
main conclusions to be extracted from Ref. \onlinecite{odr} are 
the following: (a) with reduced dimensions ODR signals are more
pronounced, and (b) experimental transition energies are 
theoretically reproduced only if valence sub-band mixing effects   
are included \cite{bauer}.

By increasing the power of the laser that creates the excitons, 
new lines corresponding to multi-excitonic transitions are
observed. In the present paper, we focus on the main biexciton
line and address the question about what are the main transitions
to be observed in an ODR experiment. As argued below, this question 
has a non-trivial answer because to many of the low-lying biexciton
states correspond very small transition probabilities, and thus 
they are very unlikely to be experimentally detected.

Following the dictates of conclusion (a) above, we study the 
biexciton in a quantum dot under the strong-confinement regime.
Very distinct biexcitonic and multiexcitonic lines have been observed in
this regime by means of confocal microscopy techniques \cite{confocal}.

Our model quantum dot is disk-shaped with a heigth, $w=8.5$ nm.
The in-plane confinement is parabolic \cite{hawrylak}, with
$\hbar\omega_0^e=7.83$ meV for electrons, leading to a characteristic
(oscillator) length of around 12 nm. Oscillator lengths
are equal for both electrons and holes. It means that 
$m_{xy}^e\omega_0^e=m_{xy}^{hh}\omega_0^{hh}$, where $m_{xy}$ are 
the in-plane masses. The Kohn-Luttinger (KL) parameters for 
GaAs of Ref. \onlinecite{bauer} are used: $\gamma_1=6.790$,
$\gamma_2 =1.924$, $\gamma_3=2.681$, $\kappa=1.2$ and
$q=0.04$. The relative dielectric constant is $\epsilon=12.5$.

Hole levels in the dot are computed from the KL hamiltonian
\cite{bauer,KL}. Instead of a numerical diagonalization, we
use second order perturbation theory to account for the 
non-diagonal terms in the KL hamiltonian. This kind of 
approximation has proven to capture many of the actual 
properties of the hole levels \cite{aprox}. Keeping the 
more relevant contributions for thin disk-shaped dots under
low magnetic fields, we get for the heavy-hole energies

\begin{equation}
E_{n\ell kj_z}^{hh}=E_{n,\ell, k,\pm 3/2}^0+
\left( \frac{8\hbar \gamma _3}
{\sqrt{3}m_0\ell_0w}\right)^2\Delta E^{\pm }, 
\label{eq1}
\end{equation}

\noindent
where the unperturbed energies are given by

\begin{eqnarray}
E_{n,\ell, k,\pm 3/2}^0&=&\frac{\hbar ^2\pi ^2}{2m_z^{hh}w^2}k^2+
\hbar \Omega^{hh}\left( 2n+|\ell |+1\right) 
\nonumber\\ &-&
\frac{\hbar \omega _c^{hh}}2\ell \pm \mu_B B g^{hh}. 
\label{eq2}
\end{eqnarray}

\noindent
$n$ and $\ell$ are radial and angular momentum projection
quantum numbers of the envolvent in-plane wave function.
$k$ labels the functions along the symmetry axis ($z$-
direction). They are taken as infinite-well functions
(assuming, for example, a dot formed from a AlGaAs-GaAs
symmetric well with high enough Al concentration). $j_z$
is the total (band) angular momentum projection along $z$.
The effective frequency $\Omega^{hh}$ is given by
$\Omega^{hh}=\sqrt{(\omega_c^{hh}/2)^2+(\omega_0^{hh})^2}$,
where $\omega_c$ is the cyclotron frequency.
$m_0$ is the electron mass in vacuum, 
$g^{hh}=3\kappa+27q/4=3.87$ and 
$\ell _0=\sqrt{\hbar/(m_{xy}^e\Omega^e)}$.
Expressions similar to Eq. (\ref{eq2}), in which masses
and frequencies are correspondingly replaced and 
$g^{lh}=\kappa+q/4=1.21$, are written 
for the light hole. The energy corrections, $\Delta E^{\pm}$, 
where the + refers to $j_z=3/2$, take the following form

\begin{eqnarray}
\Delta E_{\ell \geq 0}^{-} &=&\frac{n(1+\bar{\omega}\ell _0^2)^2}
{E_{n,\ell,1,-3/2}^0-E_{n-1,\ell+1,2,-1/2}^0}\nonumber\\
&+&\frac{(n+\ell +1)(1-\bar{\omega}\ell _0^2)^2}{E_{n,\ell,1,-3/2
}^0-E_{n,\ell +1,2,-1/2}^0} \nonumber\\
\Delta E_{\ell <0}^{-} &=&\frac{(n+|\ell |)(1+\bar{\omega}\ell _0^2)^2}
{E_{n,\ell,1,-3/2 }^0-E_{n,\ell +1,2,-1/2}^0}\nonumber\\ 
&+&\frac{(n+1)(1-\bar{\omega}\ell _0^2)^2}{E_{n,\ell,1,-3/2 }^0-
E_{n+1,\ell+1,2,-1/2}^0} \nonumber\\
\Delta E_{\ell >0}^{+} &=&\frac{(n+1)(1+\bar{\omega}\ell _0^2)^2}
{E_{n,\ell,1,3/2 }^0-E_{n+1,\ell -1,2,1/2}^0}\nonumber\\ 
&+&\frac{(n+\ell )(1-\bar{\omega}\ell _0^2)^2}{E_{n,\ell,1,3/2 }^0-
E_{n,\ell-1,2,1/2}^0} \nonumber\\
\Delta E_{\ell \leq 0}^{+} &=&\frac{(n+|\ell |+1)
(1+\bar{\omega}\ell _0^2)^2}{E_{n,\ell,1,3/2 }^0-E_{n,\ell -1,2,1/2}^0}
\nonumber\\ 
&+&\frac{n(1-\bar{\omega}\ell _0^2)^2}{E_{n,\ell,1,3/2}^0-
E_{n-1,\ell-1,2,1/2}^0}
\end{eqnarray}
where $\bar{\omega}=e B/(2\hbar)$.

The first $j_z=\pm 3/2$ hole energies obtained from an exact
diagonalization of the KL hamiltonian are drawn in Fig. \ref{fig1}
as a function of $B$. For comparison, the unperturbed  values
(\ref{eq2}) and our approximate energies (\ref{eq1}) are also drawn,
showing that band-mixing effects may lead to corrections to the
single-particle energies of around 1 meV at $B=5$ T.

Heavy-hole energies from (\ref{eq1}) are included in the biexciton
hamiltonian. For the wave function, however, we use the unperturbed
functions, neglecting hole mixing \cite{wf}. For example, for the
state with
total electron spin $S_z^e=1$, and total (band) hole 
angular momentum $J_z=3$, which is created by $\sigma^-$
polarized light, we write the biexciton spatial wave function
in the form

\begin{eqnarray}
\Psi_{1,3}&=&\frac 1 2 \sum C_{n_1 l_1, n_2 l_2; n_3 l_3, n_4 l_4}
 \times\nonumber\\
 &&\{ \phi_{n_1 l_1}(1) \phi_{n_2 l_2}(2)-\phi_{n_1 l_1}(2) 
 \phi_{n_2 l_2}(1)\}\times \nonumber\\
 &&\{ \phi_{n_3 l_3}^{3/2}(3) \phi_{n_4 l_4}^{3/2}(4)-
 \phi_{n_3 l_3}^{3/2}(4) \phi_{n_4 l_4}^{3/2}(3)\},
\label{eq4}
\end{eqnarray} 

\noindent
where indexes 1 and 2 refer to electrons, 3 and 4 to holes, and the 
sum runs over states preserving the total angular momentum
projection, $M=l_1+l_2+l_3+l_4$. Up to 16 harmonic oscillator
shells, i.e. 136 single-particle states $\phi$ for electrons, and
136 for holes, are included in (\ref{eq4}). The resulting 
large matrixes for the biexciton hamiltonian 
are diagonalized by means of a Lanczos algorithm.

The results for the first 10 transitions from the lowest 
$S_z^e=1$, $J_z=3$ state (a $M=0$ state), induced by $\sigma^{\pm}$
polarized far-infrared radiation ($\Delta M=\pm 1$) are presented
in Fig. \ref{fig2}. Levels with more than 10 \% of transition probability
are represented by solid lines, and the probability is indicated. 
Levels with probability less than 10 \% are represented by dashed
lines. We define the normalized probabilities from the expansion
coefficient of $\Psi_{final}$ in $D^{\pm} \Psi_{initial}$

\begin{equation}
prob^{\pm}=\frac{|\langle \Psi_{final}|D^{\pm}|\Psi_{initial}\rangle|^2}
 {\langle\Psi_{initial}|D^- D^+|\Psi_{initial}\rangle},
\end{equation}  

\noindent
where $D^{\pm}=r_1 e^{\pm i\theta_1}+r_2 e^{\pm i\theta_2}
-r_3 e^{\pm i\theta_3}-r_4 e^{\pm i\theta_4}$ is the dipole operator for
internal $\sigma^{\pm}$ transitions. $r,\;\theta$ are polar coordinates
in the plane.

The lowest transitions, with excitation energies below 10 meV
are interpreted as ``single-particle'' excitations.
Because of their low probabilities, they are 
unlikely to be detected in an ODR experiment.

Collective motions, in which the electron cloud oscillates in 
counterphase with respect to the hole cloud, concentrate most
of the transition probability. Their excitation energies take values
between 15 and 20 meV (the solid lines). In the biexciton, 
they are a  small-scale version of the giant-dipole resonances 
expected for larger multiexcitonic systems \cite{gdr}. Because 
of the difference between electron and hole masses and the magnetic
field, the dipole resonance splits into a few states. The
two indicated transitions at $B=0$, for example, account 
for 83 \% of the probability. At $B=5$ T, there are two states
concentrating 60 \% of the $\sigma^+$ transition probability,
and a single state accounting for 87 \% of the $\sigma^-$
probability. 

Basically, these dipole excitations are the ones to
be registered by means of ODR. Following only the solid
lines, one may notice in $\sigma^-$: (i) a ``state'' with
constant excitation energy around 16 meV which ODR signal
shall become very strong as B is increased, and (ii) a
state starting at around 20 meV which excitation 
energy rises abruptly for $B>1$ T. Whereas for $\sigma^+$
we may distinguish: (iii) a state clearly seen for $B\ge 2$
T starting at $\Delta E\approx 15$ meV and which ODR 
signal increases, (iv) a state starting
at $\Delta E\approx 17$ meV with a very strong ODR signal at
low $B$ which decreases at higher $B$ and, finally, 
(v) a state
starting at $\Delta E\approx 20$ meV which ODR signal diffuses
as $B$ is raised. The energy of all of the $M=1$ states
increases with $B$.

A careful examination of the transitions shown in
Fig. \ref{fig2} reveals a complex pattern of probability transfer
as the magnetic field is raised. For example, let us look at 
the three ``colliding'' levels at excitation energies around
18 meV for $\sigma^+$ polarization and $B=2$ T. The transition
probabilities are 42.7, 0.3 and 1.0 \% from bottom to top.
After the collision, i.e. at $B=3$ T, the probabilities become
0.0, 16.2 and 22.5 \% respectively. It means that ODR signals
may experience significant changes even with relatively small
variations of $B$. 

Similar results are presented in Fig. \ref{fig3} for the singlet
(electron), $J_z=0$ biexciton. The lowest state in this sector is also a
$M=0$ state. Low lying $M=\pm 1$ dark levels, a $\sim 15$ meV gap to the
dipole resonances, and a complex pattern of probability transfer with
increasing magnetic field are also evident in this figure.

In conclusion, we have computed the lowest lying energy levels and
transition probabilities relevant for the ODR study of the 
biexciton in a quantum dot. The states that will induce the 
most pronounced ODR signals are collective dipole excitations,
with excitation energies higher than $\hbar 
(\omega_0^e+\omega_0^{hh})$. A complex dependence of the transition
probabilities on $B$, leading to significant variations of the
position and depth of ODR signals is predicted.

The authors acknowledge support by the Committee for Research
(CODI) of the University of Antioquia and by the Caribbean Network
for Theoretical Physics. Useful discussions with B. Rodriguez are 
gratefully acknowledged.

\begin{figure}
\caption{The lowest $j_z=\pm 3/2$ states. Dashed line: Eq. (\ref{eq2}),
dotted
line: Eq. (\ref{eq1}), and solid line: exact diagonalization of the KL
hamiltonian.} 
\label{fig1}
\end{figure}

\begin{figure}
\caption{Internal transitions in the $S_{z}^e=1$, $J_z=3$ biexciton. See
explanations in the main text.}
\label{fig2}
\end{figure}

\begin{figure}
\caption{Same as Fig. \ref{fig2} for the singlet (electron),
$J_z=0$ biexciton.}
\label{fig3}
\end{figure}


\begin{references}
\bibitem{wf} For the lowest heavy hole states, the wave function almost
coincides with the unperturbed function, and orbital angular momentum can
be taken as a good quantum number. Wave function mixing becomes important
at higher excitation energies, where other effects like anharmonicity of
the lateral potential, finiteness of the barrier along $z$, etc may give
corrections of the same order to the biexciton energy.
\end{references}
\end{document}